\documentclass[prb,twocolumn,floatfix,showpacs]{revtex4}
\usepackage{graphics}% Include figure files
\usepackage{dcolumn}% Align table columns on decimal point
\usepackage{bm}% bold math
\usepackage{epsfig}
\usepackage{pictex}

\begin{document}
\title{Quantum Melting of Charge Order due to Frustration in Two-Dimensional
Quarter-Filled Systems}
\author{Jaime Merino}
\affiliation{Departamento de F\'isica Te\'orica de la Materia 
Condensada, Universidad Aut\'onoma de Madrid, Madrid 28049, Spain}
\author{Hitoshi Seo$^*$}
\affiliation{Correlated Electron Research Center (CERC), AIST Central 4, Tsukuba 305-8562, Japan}
\author{Masao Ogata}
\affiliation{Department of Physics, Faculty of Science, University of Tokyo,
Tokyo 113-0033, Japan}

\date{\today}

\begin{abstract}
The effect of geometrical frustration in a two-dimensional 
1/4-filled strongly correlated electron system is studied theoretically,
motivated by layered organic molecular crystals. 
An extended Hubbard model on the square lattice is considered, 
with competing nearest neighbor Coulomb interaction, $V$, and that of next-nearest neighbor 
along one of the diagonals, $V'$, which favor different charge ordered states. 
Based on exact diagonalization calculations, 
we find a metallic phase stabilized over a 
broad window at $V' \sim V$ even for large Coulomb repulsion strengths as 
a result of frustrating the charge ordered states. 
Slightly modifying the lattice geometry relevant to the actual organic 
compounds does not alter the results, 
suggesting that this `quantum melting' of charge order is a robust feature of frustrated strongly correlated 
1/4-filled systems. 
\end{abstract} 
\maketitle
%\sloppy

%%% Introduction
Effects of geometrical frustration in 
strongly correlated systems have been studied actively, 
where novel exotic states may arise as a result of 
competition between different `conventional' ordered phases. 
Such possibilities have been explored extensively 
in spin systems, for example, in spin-1/2 
models with frustrated antiferromagnetic (AF) 
exchange interactions of nearest neighbor (NN) 
and of next-nearest neighbor (NNN).~\cite{Diep} 
There, conventional N\'eel states can be destabilized due 
to the competition between different AF patterns and 
quantum fluctuations lead to exotic states such as spin liquids. 
In contrast, works devoted to the charge degrees of freedom 
are limited and little is known on such situations. 
One can consider an analogy between 
the classical spin systems, i.e., Ising systems, and the 
charge ordering (CO) systems.~\cite{Anderson}
However, when CO melts due to quantum fluctuation, nature of this 
transition will be different from spin systems because of Fermi 
statistics. 
Furthermore, quantum melting of CO is of particular interest 
since CO states have now been found in various materials such as 
transition metal oxides~\cite{tmo} and organic conductors,~\cite{organics} 
including systems susceptible to frustration such as in 
a triangular lattice system Na$_x$CoO$_2$ recently 
attracting interest.~\cite{Foo}

Our motivation here is CO systems having frustrated lattice structures, 
namely, quasi two-dimensional (2D) organic conductors 
$\theta$- and $\alpha$-ET$_2X$ (ET~=~BEDT-TTF, $X$:~monovalent anion).
These systems have recently been studied extensively, 
displaying a subtle competition between CO, metallic 
and superconducting phases.~\cite{HMori,McKenzie,Tajima}
Their electronic states have been studied theoretically by using
2D extended Hubbard models which include Coulomb repulsions 
of not only on-site, $U$, but also intersite, $V_{ij}$.
The latter plays a crucial role for CO 
at 3/4-filling (or equivalently 1/4-filling of holes).~\cite{Kino,Seo1}
So far, this kind of models have been studied for simple lattice 
structures such as one-dimensional (1D) chain,~\cite{Mila} 
ladder,~\cite{Vojta} 2D square lattice,~\cite{Ohta,Calandra} and so on. 
In all of these works, Wigner crystal-like CO states become 
stabilized, in general, due to the effect of $V_{ij}$ 
between NN sites when $U$ is large.

Recently, however, there appear experiments in these organic 
materials which are difficult to interpret 
within naive understandings of such CO phase transitions. 
In $\theta$-ET$_2$RbZn(SCN)$_4$, when CO was quenched by 
rapid cooling, a glassy CO state is found.~\cite{HMori,Miyagawa}
Even above the transition temperature where the system shows metallic 
resistivity,~\cite{HMori,Tajima}
the dielectric constant is unexpectedly found to 
have a frequency-dependence characteristic of insulators 
in members of $\theta$-(ET)$_2X$,~\cite{Inagaki}
while an extremely slow dynamics of CO is found in $^{13}$C-NMR in 
the above $\theta$-type compound as well as in 
$\alpha$-ET$_2$I$_3$.~\cite{Takahashi} 
These suggest effects of geometrical frustration in the charge 
degree of freedom, which we investigate in this paper. 

Fortunately, the modelling of organic systems is very simple and 
reliable.
The transfer integrals between the molecules, $t_{ij}$, 
can be obtained from the extended H{\"u}ckel method.~\cite{TMori}
In the $\theta$- and $\alpha$-type structures, 
the ET molecules are arranged in 2D planes which are close to 
triangular lattices. However, the network of $t_{ij}$ can be 
approximately described by a square lattice.~\cite{Kino}
On the other hand, the values of $V_{ij}$ are comparable
on every bond of this triangular lattice,~\cite{TMori} 
which introduce frustration in this system. 
In the previous calculations on 2D extended Hubbard models 
for these materials,~\cite{Seo1}
$U$ and $V_{ij}$ were treated at the mean-field level and 
only insulating phases were found for large interactions. 
However, we expect that a `quantum melting' of CO takes place 
due to frustration, and that a metallic region is stabilized 
even in the strongly correlated regime.
Actually, in a frustrated 1D model, 
destabilization of CO state due to quantum fluctuation 
was shown numerically.~\cite{Seo2} 
Therefore it is of fundamental interest to proceed beyond 
mean-field in those 2D models. 

%%% the model 
To study this problem, 
we first consider a simplified model for the 2D organic systems 
based on the considerations above, i.e., 
the 1/4-filled extended Hubbard model on a square lattice with NN and
NNN Coulomb repulsions: 
\begin{eqnarray}
H = &-t& \sum_{\langle ij \rangle,\sigma} (c^\dagger_{i \sigma} c_{j \sigma} +
c^\dagger_{j \sigma} c_{i \sigma}) + U \sum_{i} n_{i\uparrow}
n_{i\downarrow}
\nonumber \\
&+& V \sum_{\langle ij \rangle} n_i n_j + V' \sum_{\langle ij \rangle'} n_i n_j .
\label{ham} 
\end{eqnarray}
Here, $c^\dagger_{i \sigma}$ and ($c_{i \sigma}$) creates (eliminates) an electron of spin $\sigma$
at site $i$, $n_{i \sigma}=c^\dagger_{i \sigma}c_{i \sigma}$ and $n_i=n_{i \uparrow} + n_{i \downarrow}$. 
$t$ and $V$ are the transfer integrals and the NN Coulomb repulsion 
along the bonds of the square lattice $\langle ij \rangle$, respectively, 
and $V'$ is the NNN Coulomb repulsion along one of the diagonals $\langle ij \rangle'$. 

Some insight of the CO states in the above model can be attained by considering
the following two limits for $U/t \gg 1$, where doubly-occupied sites are
suppressed: 

{\it Case A. $V/V' \gg 1$}: An insulating state due to 
a checkerboard pattern of CO with alternating charge in every other
site (wave vector ${\bf q}=(\pi,\pi)$) is realized. 
The effective model for the spin degrees of freedom on the sites 
occupied by an electron is 
a 2D spin-1/2 AF Heisenberg model with NN $J$ and 
NNN $J'$ along both the diagonals. 
The estimated exchange couplings from fourth order perturbation in $t$ 
show $J \gg J'$ and then ${\bf q}=(\pi/2,\pi/2)$ spin ordering is expected.~\cite{McKenzie,Ohta}

{\it Case B. $V'/V \gg 1$}: The system is insulating
but with ${\bf q}=(0,\pi)$ CO forming rows or columns of occupied sites
alternating with empty ones in a stripe-type manner.  
The spin degrees of freedom is described by a quasi-1D AF Heisenberg model
with $J$ along the chains 
coupled through the interchain interaction $J'$. 
Therefore the AF 
spin ordering of ${\bf q}=(\pi,\pi/2)$ is expected. 

%%% calculation 
When $V'/V \sim 1$, the frustration is expected to 
play a major role.  
Actually, if $t=0$ and $U$ is large enough, the model is 
equivalent to the 2D Ising model with couplings $V/2$ and $V'/2$ 
as can be easily seen from Eq.~(\ref{ham}). 
In this case the ground state for exactly $V=V'$ is 
always gapless which corresponds to the metallic state 
in the present model. 
Below we will analyze Eq.~(\ref{ham}) by Lanczos exact diagonalization 
on an $L=16$ cluster 
to take account of quantum fluctuation. 

In order to determine whether the system is insulating or not 
we have calculated the Drude weight~\cite{noteED}:
\begin{equation}
\frac{D}{2 \pi e^2} = - \frac {\langle 0|T|0\rangle}{4 L}  - \frac{1}
{L} \sum_{n \ne 0} \frac{|\langle n|j_x|0\rangle|^2}{E_n-E_0}, 
\label{drude}
\end{equation}
where $E_0$ and $E_n$ are the ground state and excited state
energies of the system, respectively. 
$T$ is the kinetic energy operator 
(first term in Eq.~(\ref{ham})), $j_x$ the current operator in the $x$-direction
and $e$ the electron charge.
To explore the electronic states 
we have also calculated the charge and spin structure factors,  
$ C({\bf q})={ 1 \over L} \sum_{ij} e^{i {\bf q} \cdot {\bf R}_{ij} }\langle
n_i n_j \rangle$ and $ S({\bf q})={ 1 \over L} \sum_{ij} e^{i {\bf q} \cdot {\bf R}_{ij} }\langle
S^z_i S^z_j \rangle$, respectively, where 
${\bf R}_{ij}$ is the vector connecting two different sites 
and $S^z_i$ is the $z$-component of the spin operator. 

\begin{figure}
\begin{center}
\epsfig{file=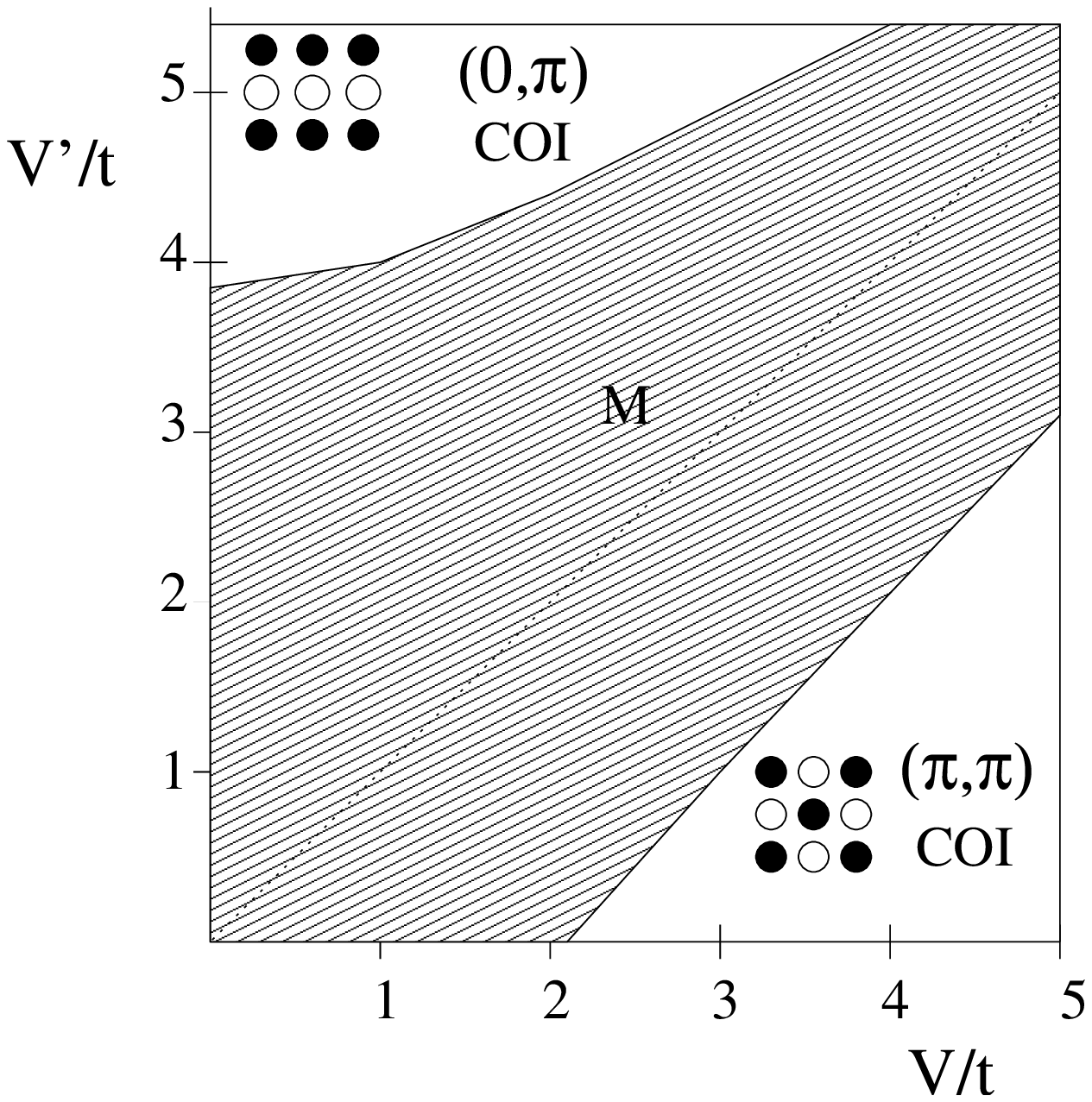,width=6cm,angle=0}
\vspace{-0.5cm}
\end{center}
\caption{Ground state phase diagram of the 1/4-filling extended Hubbard model Eq.~(\ref{ham})
for $U=10t$. The boundaries are 
extracted from exact diagionalization calculations
of the Drude weight on $L=16$ site cluster (see Fig~\ref{fig2}). 
The label M stands for metallic,
and COI for a charge ordered insulator.}
\label{fig1}
\end{figure}

The ground state phase diagram on the $V-V'$ plane 
for a fixed $U=10t$ is shown in Fig.~\ref{fig1} 
which includes CO insulating phases and a metallic phase.  
At weak coupling with small ($V/t, V'/t$) 
a metallic phase is stabilized as expected from the large 
fraction of charge carriers available in the system. 
Fixing $V'=0$ and increasing $V$ induces a metal-insulator transition
at $V \sim 2 t$ to the checkerboard CO phase 
({\it Case A} discussed above).~\cite{Calandra}
On the other hand, 
if we fix $V=0$ and vary $V'$, a metal-insulator transition to the stripe 
CO phase occurs at $V' \sim 4t$ ({\it Case B}). 
Interestingly, the metallic phase appearing at weak coupling 
extends out along the $V'=V$ line up to strong coupling in a robust manner. 

\begin{figure}
\begin{center}
\epsfig{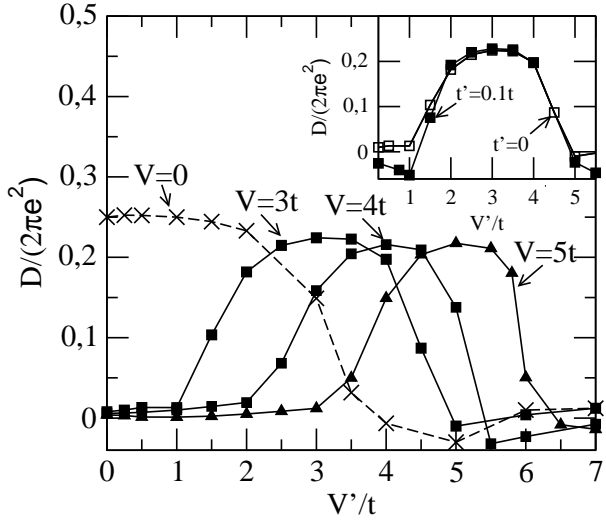} 
\vspace{-0.5cm}
\end{center}
\caption{The Drude weight calculated by exact diagonalization on $L=16$ 
for the 1/4-filled extended Hubbard model Eq.~(\ref{ham}) for $U=10t$ 
and different values of $V$. 
It is rapidly enhanced as $V'$ is increased
with $V$ fixed signalling the presence of a metallic phase.
Inset shows for $V=3t$ the variation under including a diagonal hopping, $t'=0.1t$, 
to the model, which is small.}
\label{fig2}
\end{figure}

\begin{figure}
\begin{center}
\epsfig{file=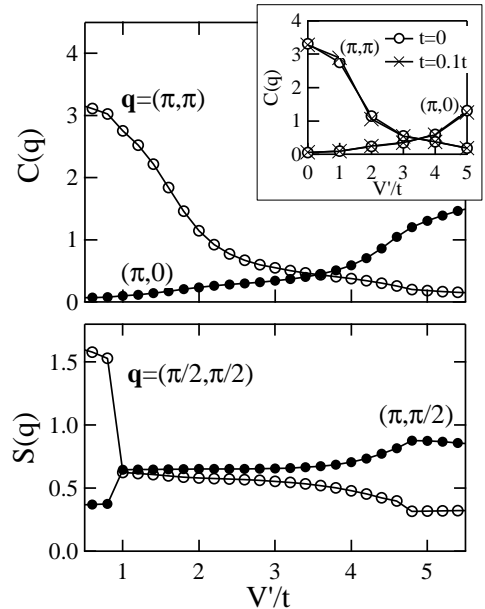,width=7.0cm,angle=0}
\vspace{-0.5cm}
\end{center}
\caption{The charge and spin structure factors, $C({\bf q})$ and $S({\bf q})$, 
for $U=10t$ and $V=3t$ 
as a function of $V'$ for wave vectors relevant for the two CO states 
discussed in the text. A crossover corresponding to the Drude weight is seen. 
Inset shows the variation of $C({\bf q})$ when small diagonal $t'=0.1t$ is 
included in the model, where the change is small. 
}
\label{fig3}
\end{figure}

In Fig.~\ref{fig2} the Drude weight 
for different values of $V$ and in Fig.~\ref{fig3} 
the structure factors for $V=3t$ as a function of $V'$, 
both for fixed $U=10t$, are plotted.  
Fixing $V$ to a large value ($V>3t$), we find that the checkerboard CO state 
with AF ordering is stabilized for small 
values of $V'$, as observed from a negligibly small Drude
weight and the large values of $C(\pi, \pi)$ and $S(\pi/2, \pi/2)$.
This insulating phase is destabilized with increasing $V'$ 
and a transition from the CO insulating phase to a metallic 
phase occurs as can be seen from the rapid increase of the Drude weight with $V'$. 
In this region, $C(\pi, \pi)$ and $S(\pi/2, \pi/2)$ are 
suppressed with $V'$ and a crossover to the stripe CO state is seen,   
characterized by the large value of $C(0,\pi)$ and $S(\pi,\pi/2)$ at large $V'$. 
When $V' \sim V$, $C({\bf q})$ becomes featureless
indicating that neither the checkerboard nor the 
stripe CO state is preferred, as expected.  

The existence of the intermediate metallic phase found here
should be robust enough in the thermodynamic limit, 
although the cluster size investigated
is limited to $L=16$.~\cite{footnote1}
Whether the CO quantum phase transition is always accompanied 
with a metal-insulator transition or not is not clear from our results. 
As a matter of fact, sudden changes of $S(\pi/2, \pi/2)$ 
and the Drude weight at $V'/t=1$ indicate a level crossing 
to a metallic state.  However, $C(\pi,\pi)$ still dominates 
over the other ${\bf q}$ values suggesting the presence of 
CO metallic states near the boundary. 
Calculations on larger systems and finite size scaling 
are neccesary to determine the CO transition lines more accurately.

This frustration induced metallic phase is found to be robust against introducing
a small hopping amplitude $t'$ along the diagonal direction $\langle ij \rangle'$,
relevant to the actual $\theta$- and $\alpha$-(ET)$_2$X materials.~\cite{HMori,Tajima} 
In the insets of Figs.~\ref{fig2} and~\ref{fig3}, comparison 
between the square lattice case Eq.~(\ref{ham}) and the 
results for the model where $t'=0.1t$ is added are shown. 
The metallic phase is stabilized in a similar $V'$ range, 
and $C(\pi,\pi)$ ($C(0,\pi)$) diplays similar 
suppression (enhancement) as $V'$ is increased. 

\begin{figure}
\begin{center}
\epsfig{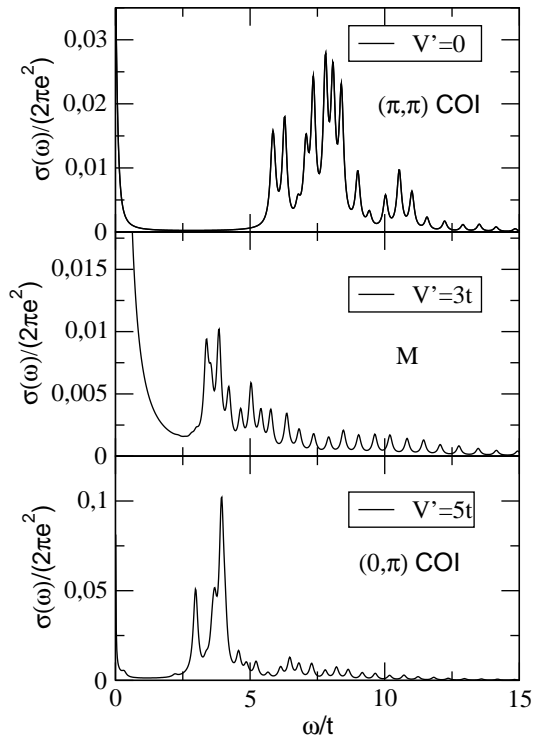}
\vspace{-0.5cm}
\end{center}
\caption{Evolution of optical conductivity $\sigma(\omega)$ of the model Eq.~(\ref{ham}) 
as the system is driven across the frustration induced metallic phase from
the checkerboard to the stripe CO phases.  Spectral weight is
transfered from high to low energies as the system is driven
through the metallic phase forming a mid-infrared band with
a feature appearing at its lower band edge and a strong Drude
component. A Lorentzian broadening of $\eta=0.1t$ is used to smoothen
out the delta peaks. 
}
\label{fig5}
\end{figure}

Finally, we discuss the evolution of the optical spectra~\cite{noteED} upon these crossovers, 
as shown in Fig.~\ref{fig5} 
for fixed $U=10t$ and $V=3t$ for different values of $V'$. 
In the case of $V'=0$ (top frame), the system is in the checkerboard CO insulating phase, 
and a gap in the spectra with a peak position centered around $8t$ is observed. 
This is naturally expected since moving a charge from the checkerboard CO state 
costs energy of $3V$ ($=9t$ in this case) in the atomic limit.~\cite{Calandra} 
On the other hand, in the stripe CO insulating phase, there are two possible 
ways of moving a charge, which are perpendicular and along the stripes, 
with the estimated energy costs of $2V'-V$ and $U-V$, respectively.
These lead to a rather broad incoherent band 
around $7t$ in the bottom frame of  Fig.~\ref{fig5},
in the case of $V'=5t$, as expected.  
Apart from this, sharp peaks at about $4t$ appear,  
which can be attributed to the 1D nature of the stripes.  
Indeed, excitonic features appear inside the Mott gap 
in the optical spectra of a 1/2-filled 1D extended Hubbard chain for $V< U/2$,~\cite{1Doptical} 
which should be an effective model for each stripe. 

In the metallic phase at intermediate $V'/V$ (see middle frame of Fig.~\ref{fig5}), a 
redistribution of the spectral weight occurs. 
Spectral weight associated with incoherent transitions due to $U$ and
$V_{ij}$ is transfered to lower energies giving rise to the Drude peak and
a mid-infrared band together with a low energy feature appearing at its lower
edge. The overall appearance of the optical conductivity is similar to the one 
observed for a 1/4-filled metal close to the checkerboard CO 
transition~\cite{Calandra} induced by $V$ ($V'=0$). 
We stress again that this is nontrivial since the interaction strengh 
is much larger here in the frustration induced case. 
However, interestingly, the mid-infrared band reaches higher energies 
making the spectrum somewhat broader while the 
feature at its lower edge appears at larger energies. 
All these `exotic' features appearing in the frustrated metallic 
phase can be attributed to the dynamics of quasiparticles 
inmersed in a charge fluctuating background and would be missed
in a mean-field analysis.
%
%%% Discussions 

In the actual organic compounds,  
$\theta$- and $\alpha$-(ET)$_2X$, $U \sim 1$ eV  and $t \sim 0.1$ eV 
which correspond to $U \sim 10t$, and $V' \sim V$ with the value of $\sim U/3$ 
is realized,~\cite{TMori} therefore the system is inherently located 
in or near to the frustration induced metallic phase. 
In reality, however, the more lower symmetry for $t_{ij}$ 
and/or the coupling to the lattice 
degree of freedom, which we do not include in our model, 
would push the system toward CO states. 
As a matter of fact, large displacements of the molecules are observed in 
$\alpha$-(ET)$_2$I$_3$ at the CO transition temperature, 
and in $\theta$-(ET)$_2X$ 
a structural change lowering the crystal symmetry takes place 
concommitantly with the CO phase transition.~\cite{organics} 
One may consider these experimental facts as suggesting that the 
CO phase is realized by relaxing the frustration effect seen in our calculations, 
since the interaction strength is already strong enough to drive the CO state. 
Another aspect worth considering when comparing our results to experimental
data is that the CO states actually observed in ET compounds have 
a different charge pattern 
as the ones discussed here,~\cite{organics}
corresponding to a `zigzag'-type pattern in our model. 
We mention that this phase is also found in our exact diagonalization study, 
e.g., for appreciably large $V'$, larger than $6t$ for $U=10t$ and $V=3t$. 
We also note that the CO pattern corresponding to the checherboard pattern has
been observed in a similar material with the $\theta$-type structure, 
i.e., $\theta$-(BDT-TTP)$_2$Cu(NCS)$_2$.~\cite{Yakushi} 
The precise determination of the phase diagram including such different CO states 
in comparison with each material, 
as well as more detailed investigation of the frustration induced metallic phase 
are left for future studies. 

In conclusion, we have found a metallic phase induced by 
frustration in 2D 1/4-filled systems which is 
robust against changing the lattice geometry from a square to
an anisotropic triangular lattice relevant to layered
organic conductors.  Our phase diagram is 
similar to that of the extended Hubbard model on the 1D zigzag chain 
structure,~\cite{Seo2} as well as to that of a frustrated 2D Bose-Hubbard Hamiltonian 
where a superfluid phase is stabilized
between two charge modulated phases.~\cite{Batrouni} 
From our findings, it is natural to conjecture that `quantum melting' of CO states 
due to geometrical frustration is rather general  
and the metallic phase near the CO phase in the actual compounds 
with geometrically frustrated structure is relevant to such situation. 
An interesting issue to address in future is whether superconductivity 
can exist between the frustration induced metallic phase and 
the CO phase mediated by charge fluctuations appearing at the phase boundary.
This kind of possible superconductivity has been studied 
in connection to organic conductors~\cite{Kobayashi} and to 
Na$_x$CoO$_2$.~\cite{Tanaka} 

\acknowledgments
J. M. acknowledges financial support from 
the Ram\'on y Cajal program from Ministerio de Ciencia y 
Tecnolog\'ia in Spain and EU under contract MERG-CT-2004-506177. 
H. S. is supported by Domestic Research Fellowship from JSPS. 
We acknowledge A. Greco, T. Hikihara, R. H. McKenzie, Y. Tanaka, and K. Tsutsui for 
discussions and comments. Part of the calculations were performed at the 
Scientific Computational Center in Universidad Aut\'onoma de Madrid.

\vspace{2mm}
\noindent
$^*$ Present address: Non-Equilibrium Dynamics Project, 
ERATO, Japan Science and Technology Agency, 
c/o KEK, Tsukuba 305-0801, Japan.


\begin{thebibliography}{99}
\bibitem{footnote1} Previous calculations~\cite{Calandra} 
at $V'=0$ show that the behavior of the Drude weight and the critical value 
for the metal-insulator transition is robust against an increase
of the size cluster from $L=16$ to $L=20$. 
Furthermore, finite size scaling of the charge gap in two-leg ladder system 
where one can reach larger sizes indicate the presence
of a metallic phase when $V' \sim V$.~\cite{Seo2}
\end{thebibliography}
\end{document}